\documentclass[aps,prd,amsmath,superscriptaddress,%
               nobibnotes,nofootinbib,preprintnumbers,showpacs,%
               onecolumn,12pt,tightenlines]{revtex4}

\usepackage{graphicx}

\usepackage{ifthen}

\newcommand{\ie}{\textit{i.e.}}
\newcommand{\eg}{\textit{e.g.}}

\newcommand{\rmd}{\ensuremath{\mathrm{d}}}

\newcommand{\erf}{\mathop{\mathrm{erf}}}

\newcommand{\gae}{%
  \ensuremath{\lower 2pt \hbox{%
    $\, \buildrel {\scriptstyle >}\over {\scriptstyle \sim}\,$}%
    }%
  }
\newcommand{\lae}{%
  \ensuremath{\lower 2pt \hbox{%
    $\, \buildrel {\scriptstyle <}\over {\scriptstyle \sim}\,$}%
    }%
  }

\newcommand{\refeqn}[2][eqn:]{Eqn.~(\ref{#1#2})}

\newcommand{\reffig}[2][fig:]{Figure~\ref{#1#2}}
\newcommand{\Reffig}[2][fig:]{Figure~\ref{#1#2}}
\newcommand{\refsec}[2][sec:]{Section~\ref{#1#2}} 

\newcommand{\ifmulticol}[2]{%
  \ifthenelse{\lengthtest{1.9\columnwidth<\textwidth}}{#1}{#2}%
}

\newcommand{\insertfig}[1]{%
    \includegraphics[keepaspectratio,width=1.00\columnwidth,
                     height=0.60\textheight]{#1}
}

\newcommand{\dRdE}{\ensuremath{\frac{\rmd\! R}{\rmd\! E}}}

\newcommand{\vmin}{\ensuremath{v_\mathrm{min}}}
\newcommand{\vmp}{\ensuremath{\overline{v}_0}}
\newcommand{\vobs}{\ensuremath{v_\mathrm{obs}}}
\newcommand{\bvobs}{\ensuremath{\mathbf{v}_\mathrm{obs}}}
\newcommand{\vesc}{\ensuremath{v_\mathrm{esc}}}
\newcommand{\Nesc}{\ensuremath{N_\mathrm{esc}}}
\newcommand{\bu}{\ensuremath{\mathbf{u}}}  
\newcommand{\bv}{\ensuremath{\mathbf{v}}}  
\newcommand{\bV}{\ensuremath{\mathbf{V}}}  

\newcommand{\fpSI}{\ensuremath{f_{\mathrm{p}}}}
\newcommand{\fnSI}{\ensuremath{f_{\mathrm{n}}}}
\newcommand{\apSD}{\ensuremath{a_{\mathrm{p}}}}
\newcommand{\anSD}{\ensuremath{a_{\mathrm{n}}}}
\newcommand{\sigmaSI}{\ensuremath{\sigma_{\mathrm{SI}}}}
\newcommand{\sigmaSD}{\ensuremath{\sigma_{\mathrm{SD}}}}

\newcommand{\sigmapSI}{\ensuremath{\sigma_{\rm p,SI}}}

\newcommand{\mhat}{\ensuremath{\hat{m}}}
\newcommand{\apSDhat}{\ensuremath{\hat{a}_{\mathrm{p}}}}
\newcommand{\anSDhat}{\ensuremath{\hat{a}_{\mathrm{n}}}}

\newcommand{\sigmapSIhat}{\ensuremath{\hat{\sigma}_{\rm p,SI}}}
\newcommand{\rhoDM}{\ensuremath{\rho_{0}}}
\newcommand{\eone}{\ensuremath{\hat{\boldsymbol{\varepsilon}}_1}}  
\newcommand{\etwo}{\ensuremath{\hat{\boldsymbol{\varepsilon}}_2}}  
\newcommand{\peone}{\ensuremath{\hat{\mathbf{e}}_1}}  
\newcommand{\petwo}{\ensuremath{\hat{\mathbf{e}}_2}}  

\newcommand{\chisqmin}{\ensuremath{\chi_{\mathrm{min}}^2}}

\begin{document}


\preprint{FTPI--MINN--09/04}


\title{Compatibility of DAMA/LIBRA dark matter detection with
       other searches in light of new Galactic rotation velocity
       measurements}

\author{Christopher Savage}
\email[]{cmsavage@physics.umn.edu}
\affiliation{
 William I.\ Fine Theoretical Physics Institute,
 School of Physics and Astronomy,
 University of Minnesota,
 Minneapolis, MN 55455, USA}

\author{Katherine Freese}
\email[]{ktfreese@umich.edu}
\affiliation{
 Michigan Center for Theoretical Physics,
 Department of Physics,
 University of Michigan,
 Ann Arbor, MI 48109, USA}

\author{Paolo Gondolo}
\email[]{paolo@physics.utah.edu}
\affiliation{
 Department of Physics,
 University of Utah,
 115 S 1400 E \#201
 Salt Lake City, UT 84112, USA}

\author{Douglas Spolyar}
\email[]{dspolyar@physics.ucsc.edu}
\affiliation{
 Department of Physics,
 University of California,
 Santa Cruz, CA, USA}

\date{\today}



\begin{abstract} 

The DAMA/NaI and DAMA/LIBRA annual modulation data, which may be
interpreted as a signal for the existence of weakly interacting dark
matter (WIMPs) in our galactic halo, are re-examined in light of new
measurements of the local velocity relative to the galactic halo.
In the vicinity of the Sun, the velocity of the Galactic disk has been
estimated to be 250~km/s rather than 220~km/s \cite{reid}. Our analysis
is performed both with and without the channeling effect included.
The best fit regions to the DAMA data are shown to move to slightly
lower WIMP masses.  Compatibility of DAMA data with null results from
other experiments (CDMS, XENON10, and CRESST~I) is investigated given
these new velocities.  A small region of spin-independent (elastic)
scattering for 7-8~GeV WIMP masses remains at 3$\sigma$.
Spin-dependent scattering off of protons is viable for 5-15~GeV WIMP
masses for direct detection experiments (but has been argued by others
to be further constrained by Super-Kamiokande due to annihilation in
the Sun).

\end{abstract} 

\maketitle


\section{\label{sec:intro} Introduction}

Among the best motivated candidates for dark matter are Weakly
Interacting Massive Particles (WIMPs).  Direct searches for dark
matter WIMPs seek to detect the scattering of WIMPs off of nuclei in
a low-background detector. 
The discovery of an annual modulation by the DAMA/NaI experiment
\cite{Bernabei:2003za}, confirmed by the new experiment
DAMA/LIBRA~\cite{Bernabei:2008yi} of the same collaboration, is the
only positive signal seen in any dark matter search.  The DAMA
collaboration has found an annual modulation in its data compatible
with the signal expected from dark matter particles bound to our
Galactic Halo \cite{Drukier:1986tm,Freese:1987wu}.  Henceforth,
we shall use the terminology ``DAMA'' to refer to all of the results to
date of the DAMA collaboration combined.  
Under the assumption of elastic scattering,
the DAMA data alone is compatible with two possible WIMP mass
ranges:  one region around
$\sim$10--15~GeV  \cite{Gelmini:2004gm,Gondolo:2005hh}
(due to scatters primarily off of Na or channeled I scatters) and
another around $\sim$60--100~GeV (due to scatters primarily off
of I).
The details of these allowed regions will be discussed further below.
However, other direct detection experiments, \eg\ 
CDMS \cite{Akerib:2003px,Ahmed:2008eu},
CoGeNT~\cite{Aalseth:2008rx},
COUPP~\cite{Behnke:2008zz},
CRESST \cite{Angloher:2002in,Stodolsky:2004dsu},
TEXONO~\cite{Lin:2007ka}, and
XENON10 \cite{Angle:2007uj,Angle:2008we}
have not found any signal from WIMPs. 
It has been difficult to reconcile a WIMP signal in DAMA with the
other negative results~\cite{previous}, for the case of canonical WIMPs
with standard weak interactions motivated by Supersymmetry (SUSY). 

An alternative way to search for WIMP dark matter is via indirect
detection of WIMP annihilation in the Sun or Earth
\cite{indirectdet:solar,indirectdet:earth}.  
The most stringent indirect bounds are from Super-Kamiokande
\cite{Desai:2004pq} and will be discussed below.
These bounds rely on additional physics, namely the strength of WIMP
annihilation to neutrinos, and thus do not apply to all WIMPs.

Previously many authors examined this question of compatibility
between all the data sets. Prior to the DAMA/LIBRA data,
Ref.~\cite{Gelmini:2004gm,Gondolo:2005hh} emphasized the low mass range
5--9~GeV; Ref.~\cite{Savage:2004fn} emphasized spin-dependent interactions; 
Ref.~\cite{Green:2003yh,Bernabei:2005hj,Bottino:2008mf} studies deviations from the
standard halo assumptions;
Ref.~\cite{Gelmini:2004gm,Gondolo:2005hh} examined nonstandard
velocity distributions (dark matter streams);
Ref.~\cite{TuckerSmith:2001hy} proposed inelastic WIMP/nucleon scattering.
In the past year DAMA/LIBRA reported new experimental results, including
presenting the modulation data in 36 separate energy bins 
and taking into account the possibility of ion channeling.

Many authors studied the new DAMA data together with the null results
of other experiments, and included the effects of ion channeling:
\cite{Bottino:2008mf,Savage:2008er,Petriello:2008jj,Chang:2008gd,
Chang:2008xa,Fairbairn:2008gz,Hooper:2008cf,Andreas:2008xy}
(although other non-WIMP candidates were studied as
well, such as mirror~\cite{Foot:2008nw}, composite~\cite{Khlopov}, and
WIMPless~\cite{Feng:2008dz} dark matter.
Non-neutralino WIMPs and Higgs exchange are examined in
Ref.~\cite{Andreas:2008xy}).
In our previous work \cite{Savage:2008er}, we delineated the remaining
regions of parameter space using the velocity information about our
motion relative to the Galactic Halo available at that time.

Recent work reexamined the local disk rotational velocities.
They examined star-forming regions 
\cite{Rygl:2008bq,VLBAGroup,Honma:2007gy,Hachisuka:2005xr}
and found that the fit to their data is significantly improved if the
local velocity is taken to be
higher than previously thought \cite{reid}.  This shift
in the velocity has the potential to change the interpretation of the
DAMA modulation signal as well as signals from other experiments. 
It is the purpose of this paper to examine this effect. 
We restrict our studies to the standard isothermal model for
the dark matter halo.  We  investigate both spin-independent
and spin-dependent coupling of WIMP/nucleus elastic scattering, 
with and without channeling included.
First, we find the new ``best fit'' regions for the DAMA data.  Then,
we compare with null results from other experiments via 
the likelihood method with goodness-of-fit. 

We begin by reviewing dark matter detection, including the role of the
new velocity information, in \refsec{Detection}.
In \refsec{Experiments} we review the data from DAMA and other
experiments.  We do not present much detail here, as we are using the
same data and analysis techniques that we discussed in our previous
paper \cite{Savage:2008er}.
In \refsec{Results} we apply our analysis techniques to obtain our
results, which can primarily be found in our figures.

\section{\label{sec:Detection} Dark Matter Detection}

WIMP direct detection experiments seek to measure the energy deposited
when a WIMP interacts with a nucleus in the detector \cite{Goodman:1984dc}.
If a WIMP of mass $m$ scatters elastically from a nucleus of mass $M$,
it will deposit a recoil energy $E = (\mu^2v^2/M)(1-\cos\theta)$,
where $\mu \equiv m M/ (m + M)$ is the reduced mass, $v$ is the speed
of the WIMP relative to the nucleus, and $\theta$ is the scattering
angle in the center of mass frame.  The differential recoil rate per
unit detector mass for a WIMP mass $m$, typically given in units of
cpd/kg/keV (where cpd is counts per day), can be written as:
\begin{equation} \label{eqn:dRdE}
  \dRdE = \frac{\sigma(q)}{2 m \mu^2}\, \rho\, \eta(E,t)
\end{equation}
where $q = \sqrt{2 M E}$ is the nucleus recoil momentum, $\sigma(q)$ is
the WIMP-nucleus cross-section, $\rho$ is the local WIMP density,
and information about the WIMP velocity distribution is encoded into the
mean inverse speed $\eta(E,t)$,
\begin{equation} \label{eqn:eta}  
  \eta(E,t) = \int_{u > \vmin} \frac{f(\bu,t)}{u} \, \rmd^3u \, .
\end{equation}
Here 
\begin{equation} \label{eqn:vmin}
  \vmin = \sqrt{\frac{M E}{2\mu^2}}
\end{equation}
represents the minimum WIMP velocity that can result in a recoil energy
$E$ and $f({\bf u},t)$ is the (time-dependent) distribution of WIMP
velocities ${\bf u}$ relative to the detector.

To determine the number of expected recoils for a given experiment
and WIMP mass, we integrate Eqn.~(\ref{eqn:dRdE}) over the nucleus
recoil energy to find the recoil rate $R$ per unit detector mass:
\begin{equation} \label{eqn:rateone}
  R(t) = \int_{E_{1}/Q}^{E_{2}/Q}\rmd E \,
         \epsilon(QE) \frac{\rho}{2 m \mu^2} \, \sigma(q) \, \eta(E,t) .
\end{equation}
$\epsilon(QE)$ is the (energy dependent) efficiency of the experiment,
due, \eg, to data cuts designed to reduce backgrounds.  $Q$ is the
quenching factor relating the observed energy $E_{det}$ (in some cases
referred to as the electron-equivalent energy) with the actual recoil
energy $E_{rec}$: $E_{det} = Q E_{rec}$ (explained in more detail
below).  The energy range between $E_{1}$ and $E_{2}$ is that of
\textit{observed} energies for some data bin of the detector (where
experiments often bin observed recoils by energy).  The quenching
factor $Q$ depends on the nuclear target and the characteristics of
the detector.

For detectors with multiple elements and/or isotopes, the total rate is
given by:
\begin{equation} \label{eqn:ratetot}
  R_{tot}(t) = \sum_i f_i R_i(t)
\end{equation}
where $f_i$ is the mass fraction and $R_i$ is the rate
(Eqn.~(\ref{eqn:rateone})) for element/isotope $i$.

The expected number of recoils observed by a detector is given by:
\begin{equation} \label{eqn:recoils}
  N_{rec} = M_{det} T R
\end{equation}
where $M_{det}$ is the detector mass and $T$ is the exposure time.

\subsection{\label{sec:SICS} Cross-Section}

The $\sigma(q)$ cross-section term in
Eqns.~(\ref{eqn:dRdE}) \&~(\ref{eqn:rateone}) is an effective
cross-section for scatters with a momentum exchange $q$.
The momentum exchange dependence appears in form factors that arise
from the finite size of the nucleus.
The total scattering cross-section generally has contributions from
spin-independent (SI) and spin-dependent (SD) couplings, with
\begin{equation} \label{eqn:CStot}
  \sigma = \sigmaSI + \sigmaSD ;
\end{equation}
these two cross-sections are described below.


\textit{Spin-independent (SI).}
For spin-independent WIMP interactions, we make the usual
assumption~\cite{Jungman:1995df} that the cross section $\sigma$ scales
with the square of the nucleus atomic number $A$ and is given by
\begin{equation} \label{eqn:SICS}
  \sigma = \sigma_{0} \, | F(E) |^2
\end{equation}
where $\sigma_{0}$ is the zero-momentum WIMP-nuclear cross-section and
$F(E)$ is a nuclear form factor,
normalized to $F(0) = 1$.  For purely scalar interactions,
\begin{equation} \label{eqn:scalar}
  \sigma_{0,\rm SI} = \frac{4 \mu^2}{\pi} [ Z \fpSI + (A-Z) \fnSI ]^2 \, .
\end{equation}
Here $Z$ is the number of protons, $A-Z$ is the number of neutrons,
and $\fpSI$ and $\fnSI$ are the WIMP couplings to the proton and nucleon,
respectively.  In most instances, $\fnSI \sim \fpSI$; the WIMP-nucleus
cross-section can then be given in terms of the WIMP-proton
cross-section as a result of Eqn.~(\ref{eqn:scalar}):
\begin{equation} \label{eqn:SICSproton}
  \sigma_{0,\rm SI} = \sigmapSI \left( \frac{\mu}{\mu_{\rm p}} \right)^2 A^2
\end{equation}
where the $\mu_{\rm p}$ is the proton-WIMP reduced mass, and $A$ is
the atomic mass of the target nucleus.
In this model, for a given WIMP mass, $\sigmapSI$ is the only free
parameter.
For the nuclear form factor we use the conventional Helmi
form~\cite{Jungman:1995df,SmithLewin,Duda:2006uk}.


\textit{Spin-dependent (SD).}
The generic form for the spin-dependent WIMP-nucleus cross-section
includes two couplings \cite{Engel:1991wq}, the WIMP-proton coupling
$a_p$ and the WIMP-neutron coupling $a_n$,
\begin{eqnarray} \label{eqn:SDCS}
  \sigma_{SD}(q) =
    \frac{32 \mu^2 G_F^2}{2 J + 1}
        \left[a_p^2 S_{pp}(q) + a_p a_n S_{pn}(q) 
      + a_n^2 S_{nn}(q) \right] .
\end{eqnarray}
Here, the quantities $a_p$ and $a_n$ are actually the axial
four-fermion WIMP-nucleon couplings in units of $2\sqrt{2} G_F$
\cite{Gondolo:1996qw,Tovey:2000mm,Gondolo:2004sc}.
The nuclear structure functions $S_{pp}(q)$, $S_{nn}(q)$, and
$S_{pn}(q)$ are functions of the exchange momentum $q$ and are
specific to each nucleus.  We take the structure functions for
Aluminum from Ref.~\cite{Engel:1995gw}; for Sodium, Iodine, and
Xenon from Ref.~\cite{Ressell:1997kx}; and for Silicon and Germanium
from Ref.~\cite{Ressell:1993qm}.  These and additional structure
functions may be found in the review of Ref.~\cite{Bednyakov:2006ux}.

\subsection{\label{sec:VelocityDist} Velocity Distribution}

We assume the Standard Halo Model (SHM) \cite{Freese:1987wu}, in which
the Galactic Halo is a simple non-rotating isothermal sphere.
We use a Maxwellian distribution, characterized
by an rms velocity dispersion $\sigma_v$, to describe the WIMP speeds,
and we will allow for the distribution to be truncated at some escape
velocity $\vesc$,
\begin{equation} \label{eqn:Maxwellian}
  \widetilde{f}(\bv) =
    \begin{cases}
      \frac{1}{\Nesc} \left( \frac{3}{2 \pi \sigma_v^2} \right)^{3/2}
        \, e^{-3\bv^2\!/2\sigma_v^2} , 
        & \textrm{for} \,\, |\bv| < \vesc  \\
      0 , & \textrm{otherwise}.
    \end{cases}
\end{equation}
Here
\begin{equation} \label{eqn:Nesc}
  \Nesc = \erf(z) - 2 z \exp(-z^2) / \pi^{1/2} ,   
\end{equation}
with $z \equiv \vesc/\vmp$, is a normalization factor.  The most
probable speed,
\begin{equation} \label{eqn:vmp}
  \vmp = \sqrt{2/3} \, \sigma_v ,
\end{equation}
is used to generate unitless parameters such as $z$.
For distributions without an escape velocity ($\vesc \to \infty$),
$\Nesc = 1$.

The WIMP component (halo or stream) often exhibits a bulk motion
relative to us, so that
\begin{equation} \label{eqn:vdist}
  f(\bu) = \widetilde{f}(\bvobs + \bu) \, ,
\end{equation}
where $\bvobs$ is the motion of the observer relative to the rest frame
of the WIMP component described by \refeqn{Maxwellian}; this motion will
be discussed below.  For the truncated Maxwellian velocity
distribution in \refeqn{Maxwellian}, the mean inverse speed,
\refeqn{eta}, becomes
\begin{equation} \label{eqn:eta2}
  \eta(E,t) =
    \begin{cases}
      \frac{1}{\vmp y} \, ,
        & \textrm{for} \,\, z<y, \, x<|y\!-\!z| \\
      \frac{1}{2 \Nesc \vmp y}
        \left[
          \erf(x\!+\!y) - \erf(x\!-\!y) - \frac{4}{\sqrt{\pi}} y e^{-z^2}
        \right] \, ,
        & \textrm{for} \,\, z>y, \, x<|y\!-\!z| \\
      \frac{1}{2 \Nesc \vmp y}
        \left[
          \erf(z) - \erf(x\!-\!y) - \frac{2}{\sqrt{\pi}} (y\!+\!z\!-\!x) e^{-z^2}
        \right] \, ,
        & \textrm{for} \,\, |y\!-\!z|<x<y\!+\!z \\
      0 \, ,
        & \textrm{for} \,\, y\!+\!z<x
    \end{cases}
\end{equation}
where 
\begin{equation} \label{eqn:xyz}
  x \equiv \vmin/\vmp \, , \quad
  y \equiv \vobs/\vmp \, , \quad \textrm{and} \quad
  z \equiv \vesc/\vmp \, ;
\end{equation}
recall $\vmin$ is given by \refeqn{vmin}.
Here, we use the common notational convention of
representing 3-vectors in bold and the magnitude of a vector in the
non-bold equivalent, \eg\ $\vobs \equiv |\bvobs|$.

Due to the motion of the Earth around the Sun, $\bvobs$ is time
dependent: $\bvobs = \bvobs(t)$.  We write this in terms of the Earth's
velocity $\bV_\oplus$ relative to the Sun as
\begin{equation} \label{eqn:vobs}
  \bvobs(t) = \bv_\odot
              + V_\oplus \left[
                  \eone \cos{\omega(t-t_1)} + \etwo \sin{\omega(t-t_1)}
                \right] \, ,
\end{equation}
where $\omega = 2\pi$/year, $\bv_\odot$ is the Sun's motion relative to
the WIMP component's rest frame, $V_\oplus = 29.8$ km/s is the Earth's
orbital speed, and $\eone$ and $\etwo$ are the directions of the Earth's
velocity at times $t_1$ and $t_1+0.25$ years, respectively.
With this form, we have neglected the ellipticity of the Earth's orbit,
although the ellipticity is small and, if accounted for, would give only
negligible changes in the results of this paper (see
Refs.~\cite{Green:2003yh,SmithLewin} for more detailed expressions).
For clarity, we have used explicit velocity vectors rather than the
position vectors $\peone$ and $\petwo$ used in
Refs.~\cite{Gelmini:2000dm,Freese:2003tt} and elsewhere (the position
vectors are more easily generalized to an elliptical orbit); the two
bases are related by $\eone = -\petwo$ and $\etwo = \peone$.

In Galactic coordinates, where $\hat{\mathbf{x}}$ is the direction to
the Galactic Center, $\hat{\mathbf{y}}$ the direction of disk rotation,
and $\hat{\mathbf{z}}$ the North Galactic Pole,
\begin{eqnarray}
  \label{eqn:eone}
    \eone &=& (0.9931, 0.1170, -0.01032) \, , \\
  \label{eqn:etwo}
    \etwo &=& (-0.0670, 0.4927, -0.8676) \, ,
\end{eqnarray}
where we have taken $\eone$ and $\etwo$ to be the direction of the
Earth's motion at the Spring equinox (March 21, or $t_1$) and Summer
solstice (June 21), respectively.

\subsection{New Measurement: Local Standard of Rest}

Unlike the Galactic disk (and the Sun), the halo has essentially
no rotation; the motion of the Sun relative to this stationary halo is
\begin{equation} \label{eqn:vsunSHM}
  \bv_{\odot,\mathrm{SHM}} = \bv_{\mathrm{LSR}} + \bv_{\odot,\mathrm{pec}}
    \, ,
\end{equation}
where $\bv_{\mathrm{LSR}}$  is the motion of the
Local Standard of Rest and $\bv_{\odot,\mathrm{pec}} = (10,13,7)$ km/s
is the Sun's peculiar velocity (\ie\ the so-called Solar Motion).
Previously, the value for the Local
Standard of Rest was taken to be $\bv_{\mathrm{LSR}} = (0,220,0)$ km/s.
Now, however, new measurements have been made
\cite{Rygl:2008bq,VLBAGroup,Honma:2007gy,Hachisuka:2005xr}
and the speed that provides the best fit to data has been found to be
higher: $\bv_{\mathrm{LSR}} = (0,250,0)$ km/s. This change in velocity
leads to the effects examined in this paper.  For an isothermal halo,
the most probable speed in \refeqn{vmp} is the same as the circular
velocity, \ie,  $\vmp$ = 250~km/s; thus the velocity dispersion is
taken to be $\sigma_v$ = 306~km/s.  The best fit for the Sun's peculiar
velocity may also shift slightly, but the changes due to this effect
are much smaller and we will ignore them.  Though the escape velocity
from the Milky Way  may also change (and is the subject of a future
work), the uncertainties are so great that we continue to use the
standard value $\vesc$ = 650~km/s.  We truncate the distribution in
\refeqn{Maxwellian} at this value.

\subsection{\label{sec:Modulation} Annual Modulation}

The count rate in WIMP detectors will experience
an annual modulation as a result of the motion of the Earth around the
Sun described above \cite{Drukier:1986tm,Freese:1987wu}.  Typically
the count rate (\refeqn{dRdE}) has an approximate
time dependence
\begin{equation} \label{eqn:dRdEapprox}
  \dRdE(E,t) \approx S_0(E) + S_m(E) \cos{\omega(t-t_c)} ,
\end{equation}
where $t_c$ is the time of year at which $\vobs(t)$ is at its maximum.
$S_0(E)$ is the average differential recoil rate over a year and
$S_m(E)$ is referred to as the modulation amplitude (which may, in fact,
be negative).  The above equation is a reasonable approximation for the
SHM we are considering in this paper, but is not valid for all halo
models, particularly at some recoil energies for dark matter streams;
see Ref.~\cite{Savage:2006qr} for a discussion.  For the SHM,
\begin{equation} \label{eqn:SmSHM}
  S_m(E) = \frac{1}{2} \left[
             \dRdE(E,\,\textrm{June 1}) - \dRdE(E,\,\textrm{Dec 1})
           \right] .
\end{equation}
Experiments such as DAMA will often give the average amplitude over
some interval,
\begin{equation} \label{eqn:Sm}
  S_m = \frac{1}{E_2 - E_1} \int_{E_1}^{E_2} \rmd E \, S_m(E) .
\end{equation}
The Earth's speed relative to the
halo, $\vobs(t)$, is maximized around June 1.  While this date varies
with changes in the velocity distribution, the change is negligible
(less than a day) for the values we are considering.

\subsection{\label{sec:Parameters} Parameter Space}

Many of the parameters that factor into the expected recoil rates for a
scattering detector are unknown, including the WIMP mass, four
WIMP-nucleon couplings (SI and SD couplings to each of protons and
neutrons), the local WIMP density, and the WIMP velocity distribution
in the halo.  In this paper, we shall fix the halo model to the SHM and
take the local dark matter
density $\rhoDM$ to be the estimated average density in the
local neighborhood, 0.3~GeV/cm$^3$.
In addition, we shall take
$\fpSI = \fnSI$ (equal SI couplings) so that there are only three
independent scattering couplings; the SI coupling will be given in
terms of the SI scattering cross-section off the proton, $\sigmapSI$.
The parameter space we examine will then consist of the four parameters
$m$, $\sigmapSI$, $\apSD$, and $\anSD$.

\section{\label{sec:Experiments} Experiments}

\subsection{\label{sec:DAMA} DAMA}

The DAMA experiments remain the only direct detection experiments to
observe a signal.  
We primarily focus on the annual modulation signal found by the DAMA
group.  Details of our analysis can be found in our previous paper
\cite{Savage:2008er}.

\textit{Ion Channeling (IC):} 
Here we review the experimental effect of ion channeling, which has been
found to affect the results.  In general only a fraction (known as
the quenching factor) of the recoil energy deposited by a WIMP
is transferred to electrons and is converted into useful signal in
the DAMA detector (\eg\ ionization or scintillation); the remainder is
converted into phonons and heat and goes undetected.  Hence measured
energies must be corrected for this behavior to obtain the proper recoil
energy, by dividing by this quenching factor.  
The DAMA detector is
composed of NaI, with quenching factors $Q_{Na} = 0.3$ and $Q_I = 0.09$.
Yet, recently, the DAMA collaboration pointed out that some fraction of
the nuclei may recoil along a crystal axis or plane
\cite{Bernabei:2007hw}, in which cases the recoiling nucleus can travel
relatively large distances along the ``channels'' between the crystal
atoms
(as opposed to the typical case, where the recoiling nucleus quickly
collides with nearby atoms).
Such recoiling nuclei lose very little energy to other atoms and to
heat, giving nearly all their energy to electrons.  For these cases the
measured energy corresponds very nearly to the energy deposited by the
WIMP and thus have quenching factor $Q \approx 1$.  As a consequence,
the detector is sensitive to lower mass WIMPs than previously thought.
The new channeling studies revived
the possiblity of DAMA's compatibility with other data sets,
particularly at low masses.
We shall examine the DAMA results both with and without including the
IC effect. Details of our analysis can be found in our previous paper
\cite{Savage:2008er}.


\textit{Total Rate:}
While we mainly examine the DAMA experiment by analyzing the modulation
amplitude $S_m$, DAMA can additionally constrain parameter space using
the total rate ($S_0$ in \refeqn{dRdEapprox}) as shown in Figure~1 of
Ref.~\cite{Bernabei:2008yi}.
The DAMA detectors do not strongly discriminate between WIMP scatters
and background events, so an unknown and possibly large portion of the
total observed events may be due to backgrounds (the backgrounds are
presumed not to vary with time and do not contribute to the modulation).
In the same manner as the various null experiments, we use the total
number of events observed by DAMA
to constrain the parameter space by excluding regions that
would predict more events than observed.

We will show regions excluded at the
90\% level due to the total rate for DAMA.  Constraints will be shown
with and without the channeling effect.
Note that some choices of parameters that produce the correct
modulation signal may actually predict a total rate larger than
observed and, hence, those parameter choices are incompatible with
the full DAMA data set.

\subsection{\label{sec:NullResults} Null Results}

Numerous other experiments have searched for a dark matter signal, but
all these experimental results, apart from DAMA, are consistent with no
WIMP signal.  

\textit{Direct Detection:}
In our previous paper \cite{Savage:2008er} we discussed in detail our
use of data from many experiments. We do not repeat this information
here.  In this paper we restrict our analysis to the most constraining
experiments:
CDMS \cite{Ahmed:2008eu} \cite{Akerib:2003px},
XENON \cite{Angle:2007uj,Angle:2008we},
and CRESST~I \cite{Angloher:2002in}.
While COUPP \cite{Behnke:2008zz} could be very powerful,
insufficient information on the data runs is provided to
reproduce their results and extend them to arbitrary couplings.

For the various null experiments, we define constraints in parameter
space at a certain exclusion level $1-\alpha$, typically 90\%, as the
parameters for which the probability of seeing the experimental
result is $\alpha$.  Parameters outside the corresponding contours
would yield the observed result with a probability less than $\alpha$.
We say the parameters within those contours are compatible within the
$1-\alpha$ level and parameters outside are excluded at the $1-\alpha$
level.  The probabilities are determined using different statistical
values, described in our previous paper, in different cases.


\textit{Indirect Detection:}
An alternative way to search for WIMP dark matter 
is via indirect detection of WIMP annihilation in the Sun or Earth.  
When WIMPs pass through these objects
\cite{indirectdet:solar,indirectdet:earth}, a small fraction may
be captured, sink to the core, and annihilate with one another to
produce a neutrino signal observable in
Earth-based detectors such as Super-Kamiokande \cite{Desai:2004pq}
(Super-K), 
AMANDA \cite{Ahrens:2002eb,Ackermann:2005fr},
IceCube \cite{Ahrens:2002dv} and ANTARES \cite{Blanc:2003na}
\footnote{
Other indirect detection methods search for WIMPs that
annihilate in the Galactic Halo or near the Galactic Center where they
produce neutrinos, positrons, or antiprotons that may be seen in
detectors on the Earth
\cite{indirectdet:galactichalo,indirectdet:galacticcenter}.}. 
Since the Sun is mostly made of hydrogen (which has
spin), its capture of WIMPs depends primarily on the spin-dependent
WIMP/nucleon cross section which can thus be constrained.
The most stringent indirect bounds are from Super-K
\cite{Desai:2004pq,Hooper:2008cf}.   However, Super-K is not shown in the
plots because their result cannot be readily
reanalyzed for a modified halo.

The Super-K constraints rely on the following assumptions:
(1) the WIMP/anti-WIMP abundance is not highly asymmetric, which would
suppress the annihilation rate;
(2) the annihilation cross-section is sufficiently high so that the
capture rate of WIMPs in the Sun (via scattering off of nuclei in the
Sun) is in equilibrium with the annihilation rate; and
(3) the WIMP does not annihilate predominantly into the light quarks,
which do not yield neutrinos in sufficient quantities and energies to
be observed.
For the Constrained Minimal Supersymmetric Standard Model, these
assumptions are mainly satisfied for neutralinos in the parameter space
of interest.  In general, however, these assumptions need not be
satisfied, so the Super-K constraints should be applied with caution.

\subsection{Analysis Techniques}

We use two statistical tests to interpret the DAMA modulation data.
First, we find the ``best fit'' regions in WIMP parameter space to fit
the DAMA data by itself.  More specifically, we use
a likelihood ratio method with a global fit
of four parameters to find the preferred parameters to produce the DAMA
signal, and plot these in the plane of the WIMP/nucleon scattering
cross section vs.\ WIMP mass.
Second, we use
a $\chi^2$ goodness-of-fit test (g.o.f.) to indicate
which parameters are compatible with the DAMA signal.  See
Ref.~\cite{Amsler:2008zz} for a short review of statistics or
Ref.~\cite{James:2006} for more extensive discussions.

\subsubsection{Best Fit to DAMA data (Likelihood Ratio Method)}
   
To determine the most likely parameters for producing the DAMA signal,
we use the maximum likelihood method, based on the likelihood ratio
\begin{equation} \label{eqn:LR}
  \frac{L(S_{m,k}|m,\sigmapSI,\apSD,\anSD)}%
       {L(S_{m,k}|\mhat,\sigmapSIhat,\apSDhat,\anSDhat)} ,
\end{equation}
where $L$ is the likelihood function, $S_{m,k}$ are the observed
modulation amplitudes in each bin, and $\mhat$, $\sigmapSIhat$,
$\apSDhat$, and $\anSDhat$ are the values of the parameters
that maximize the likelihood for the observed $S_{m,k}$.  The
denominator of the above equation is the maximum likelihood value
$L_{\mathrm{max}}$.
Confidence regions in the parameters are determined by
\begin{equation} \label{eqn:LRCR}
  2\, \ln L(m,\sigmapSI,\apSD,\anSD)
  \ge 2\, \ln L_{\mathrm{max}} - 2\, \Delta \ln L ,
\end{equation}
where the value of $\Delta \ln L$ corresponds to the
confidence level (C.L.) of the confidence region.  Since each bin is
normally distributed, this equation may instead by given in terms of
the $\chi^2$,
\begin{equation} \label{eqn:ChiSqCR}
  \chi^2(m,\sigmapSI,\apSD,\anSD)
  \le \chisqmin + \Delta \chi^2 ,
\end{equation}
where
\begin{equation} \label{eqn:ChiSq}
  \chi^2(m,\sigmapSI,\apSD,\anSD)
    \equiv \sum_k \frac{(S_{m,k} - S_{m,k}^{\mathrm{Th}})^2}{\sigma_k^2} ,
\end{equation}
$\sigma_k$ is the uncertainty in $S_{m,k}$,
$S_{m,k}^{\mathrm{Th}} \equiv S_{m,k}^{\mathrm{Th}}(m,\sigmapSI,\apSD,\anSD)$
is the expected amplitude in a particular bin for the given parameters,
and $\chisqmin$ is the minimum value of $\chi^2$.

The $\Delta \chi^2$ limit corresponding to different C.L.'s
depends upon the number of parameters that have been minimized
and, in the large data sample limit, is determined from a $\chi^2$
distribution with the degrees of freedom (d.o.f.) equal to the number
of minimized parameters.
For the four parameters considered here, $\Delta \chi^2$ =
7.8 (90\% C.L.), 16.3 (3$\sigma$), 34.6 (5$\sigma$), and
60.3 (7$\sigma$).

The confidence region for DAMA as described here is, in fact, a
4-dimensional region in the ($m,\sigmapSI,\apSD,\anSD$) parameter
space.  We shall be showing 2-dimensional slices of this larger
region for particular cases, such as the $\apSD = \anSD = 0$ slice
corresponding to SI only scattering.  This is \textit{not} equivalent
to fixing $\apSD = \anSD = 0$ and determining a 2-dimensional
confidence region by minimizing only over ($m,\sigmapSI$).  In the
latter case, a confidence region will always be defined in a
2-dimensional plane while, in the former, there may be no slice in
that plane if that plane represents a poor ``fit'' relative to the
overall parameter space.

Note, as will be discussed in \refsec{Results}, the confidence regions
obtained via this method yield the most \textit{preferred} parameters
for producing the signal.  This method, in and of itself, does
\textit{not} imply parameters outside the confidence regions are
necessarily a bad fit to the data (or, conversely, that parameters
inside these regions are a good fit) and one should be careful using
these regions to compare with other experiments \footnote{In
    statistical parlance, the determination of the confidence
    region for this method is \textit{decoupled} from the
    goodness-of-fit.}.

\subsubsection{Goodness-of-Fit}

We use the alternative statistic of goodness-of-fit (g.o.f.) to indicate
when parameters are incompatible with DAMA data, as opposed to finding
the best fit region as we did previously.  This statistic allows for
comparison of DAMA modulation data both with the total count rate from
the DAMA data itself and with null results from other experiments
(whereas it would be inappropriate to use the best fit regions for this
purpose).  To conservatively indicate parameters that are statistically
compatible with the DAMA data set, this method indicates regions at
which the $\chi^2$ falls within
a given level using a simple $\chi^2$ goodness-of-fit (g.o.f.) test on
the data.  In contrast to the previous  analysis method, there
is no fit to the parameters here.  The g.o.f.\ regions are defined as
those parameters for which
\begin{equation} \label{eqn:GOFChiSq}
  \chi^2(m,\sigmapSI,\apSD,\anSD) \le \chi_{\mathrm{GOF}}^2 ,
\end{equation}
where $\chi_{\mathrm{GOF}}^2$ is the value at which the $\chi^2$
cumulative distribution function (CDF) for \eg\ 36 d.o.f.\ is equal to
the desired level of compatibility.  That is, for a desired
compatibility level of $1 - \alpha$, there is a probability $\alpha$
that $\chi^2$ will exceed $\chi_{\mathrm{GOF}}^2$.  Alternatively, we
can say parameters outside of the region are excluded at the
$1 - \alpha$ level.

To improve the ability of the g.o.f.\ test to exclude
some parameters, we combine all DAMA bins over 10-20~keVee into a
single bin 
(energies at which negligible signal is expected),
resulting in a total of 17 data bins used for this test.
For the 17 d.o.f.\ $\chi^2$ distribution, values of 24.8, 37.7, and 61.6
are excluded at the 90\%, 3$\sigma$, and 5$\sigma$ levels, respectively.

While the g.o.f.\ regions defined here are, in fact, confidence
regions---indicating the likely parameters to produce the DAMA signal
in our theoretical framework---we do not use these regions in that
manner.  The determination of a confidence region \textit{assumes}
that the theoretical framework is correct, \ie\ there exists some
choice of parameters that is correct.  That assumption may not be
valid if, \eg, the standard halo model is not a reasonable
approximation of the actual halo.
Instead, we will more conservatively only use this g.o.f.\ test to
\textit{exclude} parameters outside of the corresponding regions.
That is, parameters outside of the DAMA g.o.f.\ regions are
incompatible with the DAMA signal at the given level
within the context of this model.

\section{\label{sec:Results} Results}

\begin{figure}
  \insertfig{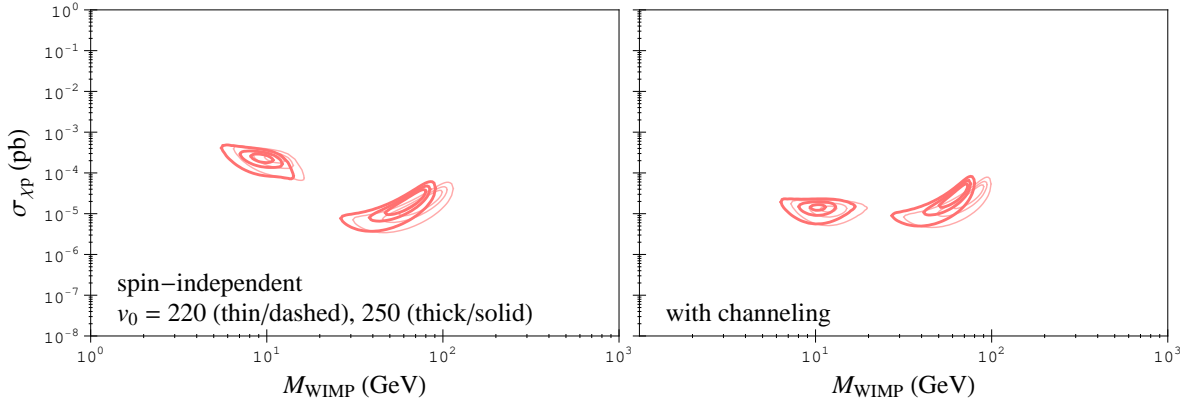}
  \caption{
    DAMA best fit parameters (cross section and WIMP mass) for SI
    only scattering.
    The DAMA contours are shown at 5$\sigma$, 3$\sigma$, and 90\% C.L.\ 
    (outermost contour is 5$\sigma$).
    Light/thin lines are for the old value of local
    standard of rest ($v_0 = 220$~km/s) while the dark/thick lines are
    the new value ($v_0 = 250$~km/s).
    The DAMA preferred regions are determined without (Panel~A) and
    with (Panel~B) the channeling effect.
    }
  \label{fig:SIpLR}
\end{figure}

\begin{figure}
  \insertfig{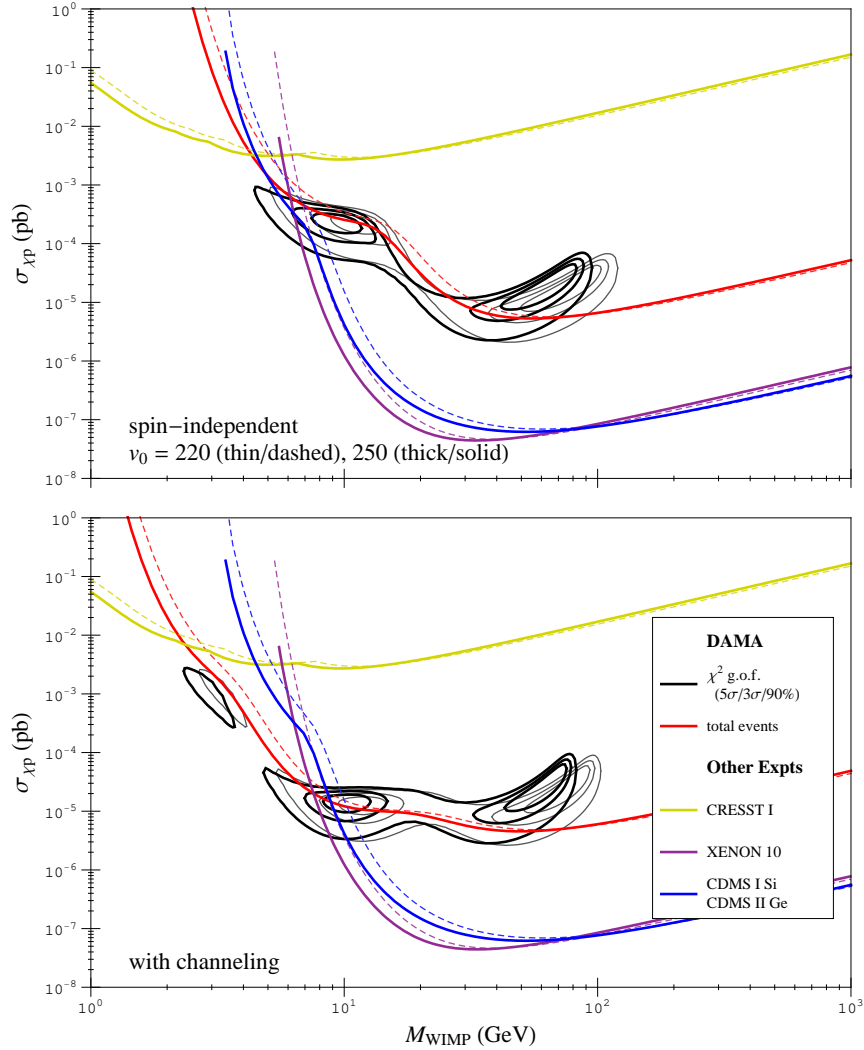}
  \caption{
    Spin-independent WIMP/nucleon scattering cross section vs.\ WIMP
    mass: comparison of DAMA with other null experiments (CRESST~I,
    XENON10, and CDMS) using the goodness-of-fit statistic.
    Thin/dashed lines are for the old value of local standard of rest
    ($v_0 = 220$~km/s) while the thick/solid lines are the new value
    ($v_0 = 250$~km/s).  The  bound from the total events from DAMA is
    also shown (red solid line).
    Panel~A does not include the effects of ion channeling while
    Panel~B does take it into account.
    The DAMA contours are shown at 5$\sigma$, 3$\sigma$, and 90\% C.L.\ 
    (outermost contour is 5$\sigma$).
    One can see that, with IC included, a sliver of parameter space
    remains viable at 3$\sigma$ for 7-8 GeV WIMPs.
    }
  \label{fig:SIpGOF}
\end{figure}

\begin{figure}
  \insertfig{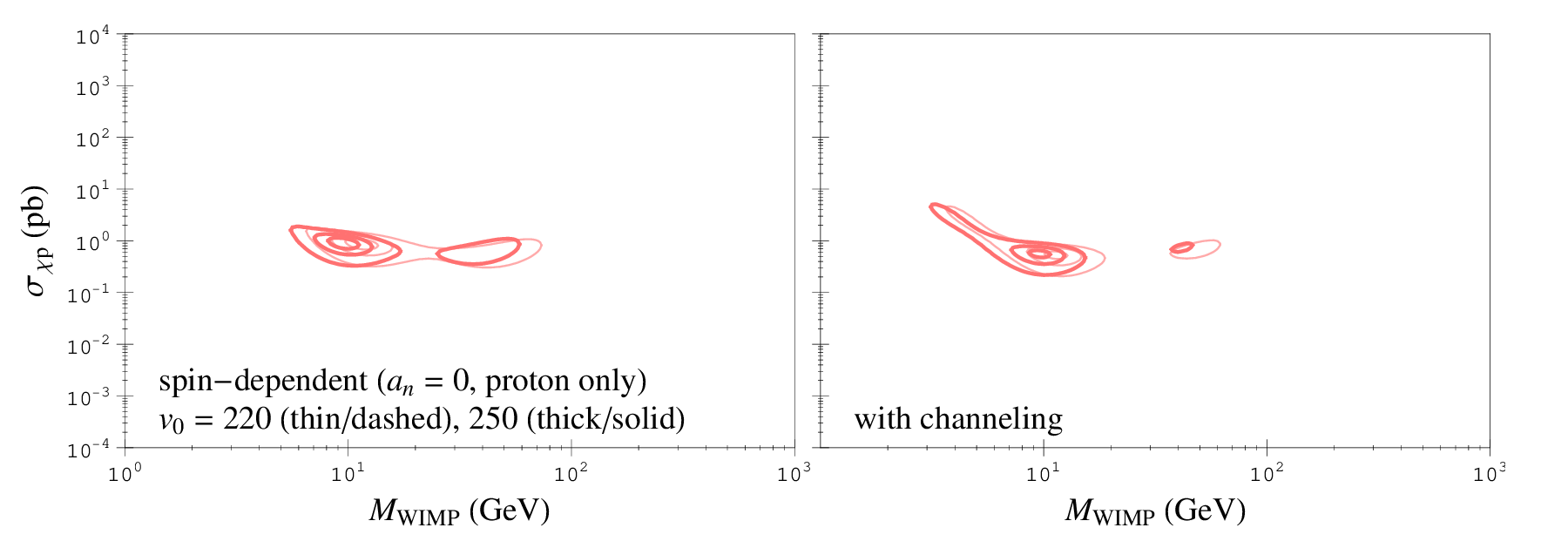}
  \caption{
    Same as \reffig{SIpLR} but for SD proton-only scattering $(a_n=0)$.
    Note there are no 3$\sigma$ or 90\% C.L.\ contours at the higher
    WIMP mass region.
    }
  \label{fig:SDpLR}
\end{figure}

The likelihood ratio analysis of DAMA yields a 4-dimensional confidence
region over the ($m,\sigmapSI,\apSD,\anSD$) parameter space.
In Figures~\ref{fig:SIpLR}, \ref{fig:SDpLR}, and~\ref{fig:SDnLR}, we
show the SI only ($\apSD = \anSD = 0$), SD proton-only
($\sigmapSI = \anSD = 0$), and SD neutron-only ($\sigmapSI = \apSD = 0$)
slices of this confidence region, respectively.
The thin/dashed curves are for the old velocity of the  local 
standard of rest $v_0=220$~km/s while the solid curves are for the new
local standard of rest $v_0 = 250$~km/s.  In all cases, the first panel
of any figure is without channeling, while
the second panel is with channeling taken into account for DAMA.

We also present results using the goodness-of-fit statistic to compare
DAMA's positive annual modulation signal with its own bound on the
total count rate plus null results from other experiments (CRESST~I,
CDMS, and XENON10) to find what regions of parameter space remain
compatible with all of these direct detection searches.  The total rate
from DAMA itself restricts the high mass end, and in some cases
rules out the entire higher mass window (near 60 GeV).

While we present our results in three particular
cases---SI only, SD proton-only, and SD neutron-only---it is important
to keep in mind that there is a larger region of parameter space that
includes mixed couplings.

\subsection{Spin-Independent Couplings (SI)}

\begin{figure}
  \insertfig{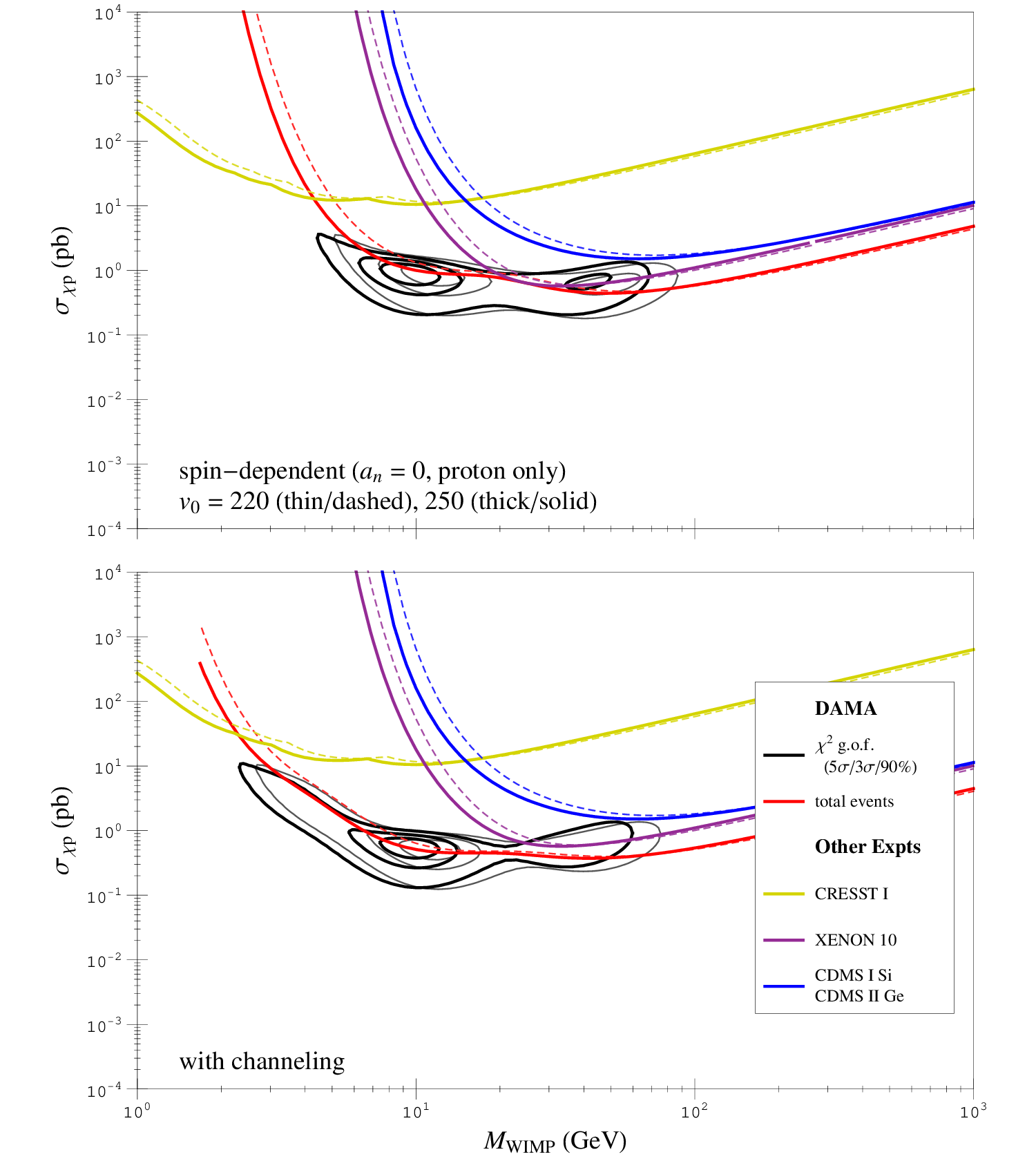}
  \caption{
    Same as \reffig{SIpGOF} but  for SD proton-only scattering $(a_n=0)$.
    Without Super-K included compatiblity is viable for 5-15 GeV WIMP
    masses for direct detection experiments (but has been argued by
    others to be further constrained by Super-Kamiokande due to
    annihilation in the Sun).
    }
  \label{fig:SDpGOF}
\end{figure}


\Reffig{SIpLR} shows the best fit regions to the DAMA data for
spin-independent couplings, both with and without channeling.  One can
see that the effect of the larger velocity is to shift the regions to
slightly lower masses.  \Reffig{SIpGOF} uses the goodness-of-fit
(g.o.f.) statistic to compare the DAMA regions with the null
experiments, for the SI couplings.  Most of the parameter space is
incompatible, with the exception of a small sliver at 7-8 GeV WIMP
masses that are still in agreement with all experiments to 3$\sigma$.

\subsection{Spin-Dependent Couplings (SD)}

\begin{figure}
  \insertfig{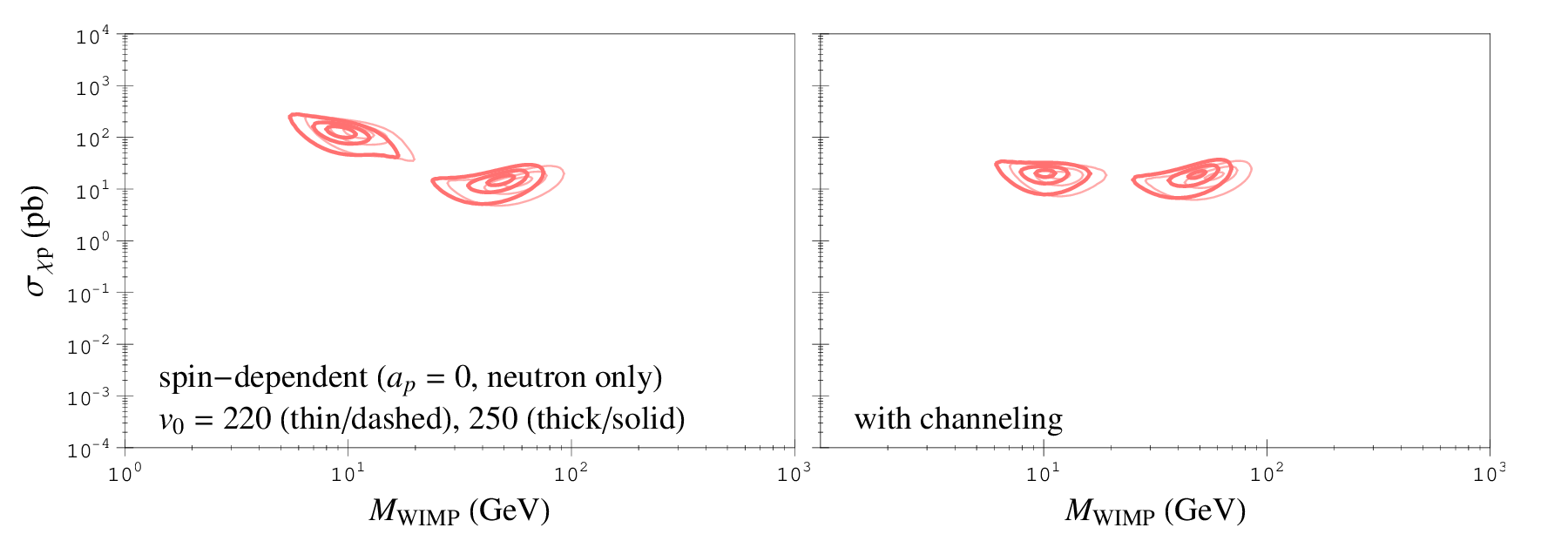}
  \caption{
    Same as \reffig{SIpLR} but for SD neutron-only scattering $(a_p=0)$.
    }
  \label{fig:SDnLR}
\end{figure}

\begin{figure}
  \insertfig{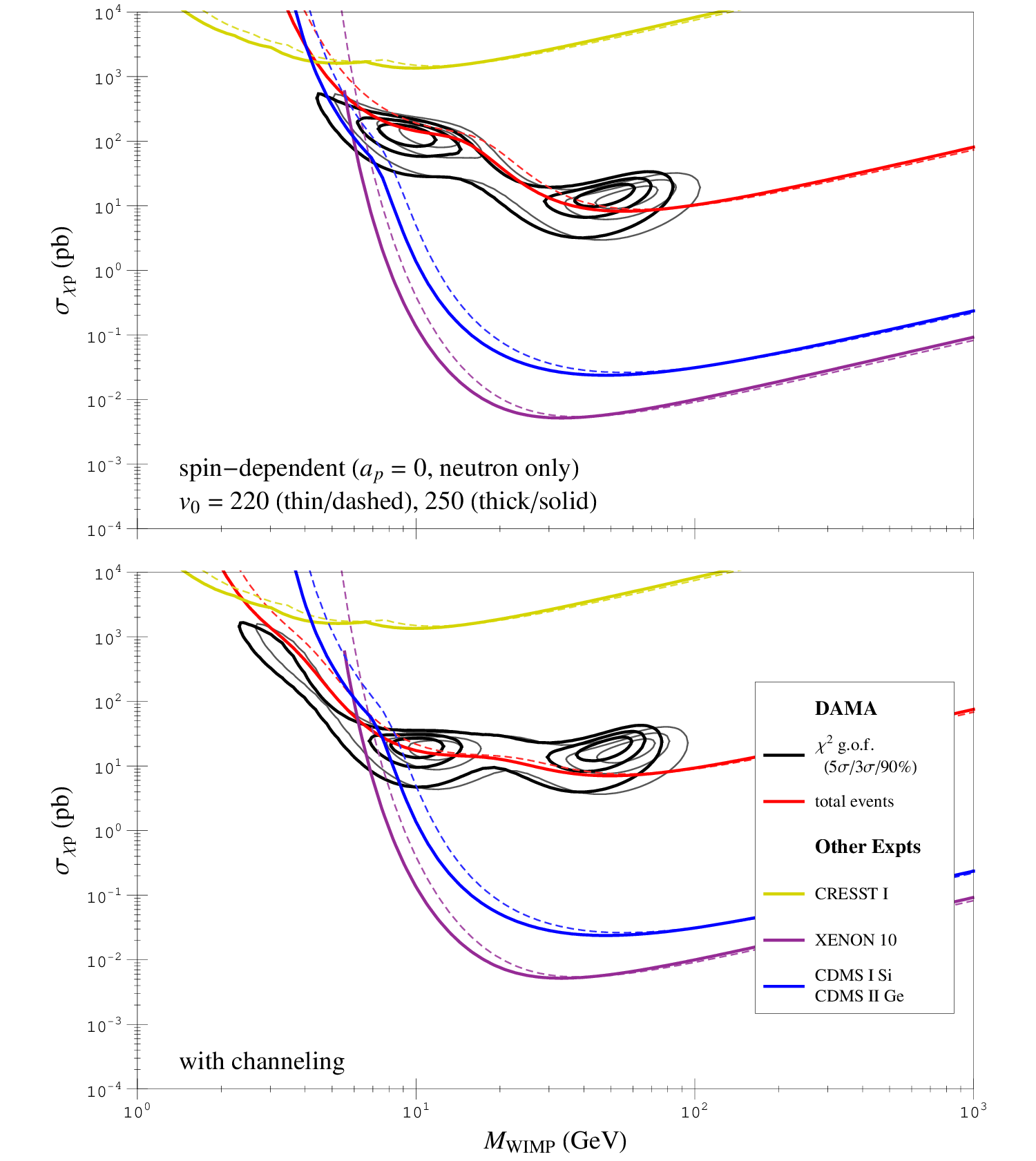}
  \caption{
    Same as \reffig{SIpGOF} but for SD neutron-only scattering $(a_p=0)$.
    No regions of parameter space remain.
    }
  \label{fig:SDnGOF}
\end{figure}

\textit{Proton only  $(a_n=0)$:}
\Reffig{SDpLR} shows the best fit regions to the DAMA data
for spin-dependent couplings for proton only $(a_n = 0$), 
both with and without channeling.  Again, the best fit regions move to
slightly lower masses.  In \reffig{SDpGOF} one can see that all WIMP
masses suitable for the DAMA data are compatible with all the other
direct detection experiments.  With channeling, the 90\% region is for
7-12 GeV WIMP masses and the 3$\sigma$ region is for 5-15 GeV WIMP
masses.  However, the COUPP results, 
once made available, can severely change these conclusions.  

The Super-K indirect detection results, which have not been plotted,
can also change the conclusions.  Ref.~\cite{Hooper:2008cf}
have studied the constraints on neutralino WIMPs with different
annihilation channels. If we use their results (which
have not been done for the new velocity measurements),
the conclusion for the case of SD proton-only couplings is that
virtually all neutralinos are incompatible with the combination of
SuperK plus all direct detection experiments; this statement
assumes that there is no WIMP/antiWIMP asymmetry as well as the
standard annihilation cross section required for thermal WIMPs. 
[Note that Super-K would rule out SD proton-only
couplings, not neutron-only or SI couplings.]

The low mass region found in the SI and the SD proton-only cases
persists with a combination of these two couplings.  That is, there
are portions of parameter space with both SI couplings and SD couplings
to the proton that are compatible with all the direct detection
experiments.

\textit{Neutron only $(a_p=0)$:}
\Reffig{SDnLR} shows the best fit DAMA regions for spin-dependent
couplings for neutron only $(a_p=0)$.  \Reffig{SDnGOF} shows that there
is no region of WIMP masses which can be compatible with all
experiments in this case.
While this SD neutron-only case is incompatible, 
non-zero SD couplings to the neutron are not in general incompatible:
cases with mixed couplings (with both $a_p \ne 0$ and $a_n \ne 0$)
can still be found that are consistent with all the experimental
results.

\section{\label{sec:Conclusion} Conclusion}

Our general result is that the new determination of the local standard
of rest moves the best fit WIMP masses to the DAMA data to lower values.   
We had hoped that the DAMA data, which is very dependent on the WIMP
velocity behavior, would change enough to allow new regions of
parameter space compatible with all experiments.  However, the bounds
from the other experiments shifted by about the same amount, so that
very little changed due to the analysis in this paper.


\begin{acknowledgments}
  We acknowledge the Michigan Center for Theoretical Physics at the
  University of Michigan, where this project came into existence while
  we were all attending the workshop on LHC and Dark Matter.
  K.F.\ acknowledges the support of the DOE  via the University of
  Michigan.
  P.G.\ was partially supported by NSF grant PHY-0756962 at the
  University of Utah.
  C.S.\ acknowledges the support of the William I.\ Fine Theoretical
  Physics Institute at the University of Minnesota.
  D.S.\ acknowledges a GAANN fellowship.
\end{acknowledgments}




\end{document}